\documentclass[prd,twocolumn,twoside,preprintnumbers,superscriptaddress,nofootinbib]{revtex4}

\usepackage{amsmath,slashed}
\usepackage{graphicx,graphics,color}
\usepackage{dcolumn}
\usepackage[hyperfootnotes=false]{hyperref}
\usepackage{xspace}
\usepackage{enumerate}
\usepackage{varwidth}

\usepackage{amssymb}
\usepackage{mathtools}
\usepackage{dsfont}
\usepackage{multirow}

\newcommand{\MeV}{\,\text{MeV}}

\newcommand{\mpi}{M_\pi}
\newcommand{\Fpi}{F_\pi}

\newcommand{\mW}{M_W}
\newcommand{\mN}{m_N}
\newcommand{\eps}{\epsilon}

\newcommand{\Order}{\mathcal{O}}

\newcommand{\diff}{\text{d}}
\newcommand{\qq}{\mathbf{q}}

\newcommand{\qg}{\mathbf{q}_\gamma}

\newcommand{\beq}{\begin{equation}}
\newcommand{\eeq}{\end{equation}}

\allowdisplaybreaks[1]

\begin{document}

\preprint{INT-PUB-24-020, LA-UR-24-25162}

\title{Radiative corrections to superallowed $\boldsymbol{\beta}$ decays in effective field theory}

\author{Vincenzo Cirigliano}
\affiliation{Institute for Nuclear Theory, University of Washington, Seattle WA 91195-1550, USA}
\author{Wouter Dekens}
\affiliation{Institute for Nuclear Theory, University of Washington, Seattle WA 91195-1550, USA}
\author{Jordy de Vries}
\affiliation{Institute for Theoretical Physics Amsterdam and Delta Institute for Theoretical
Physics,University of Amsterdam, Science Park 904, 1098 XH Amsterdam, The Netherlands}
\affiliation{Nikhef, Theory Group, Science Park 105, 1098 XG, Amsterdam, The Netherlands}
\author{\\Stefano Gandolfi}
\affiliation{Theoretical Division, Los Alamos National Laboratory, Los Alamos, NM 87545, USA}
\author{Martin Hoferichter}
\affiliation{Albert Einstein Center for Fundamental Physics, Institute for Theoretical Physics, University of Bern, Sidlerstrasse 5, 3012 Bern, Switzerland}
\author{Emanuele Mereghetti}
\affiliation{Theoretical Division, Los Alamos National Laboratory, Los Alamos, NM 87545, USA}

\begin{abstract}
 The accuracy of $V_{ud}$ determinations from superallowed $\beta$ decays critically hinges on control over radiative corrections. Recently, substantial progress has been made on the single-nucleon, universal corrections, while nucleus-dependent effects, typically parameterized by a quantity $\delta_\text{NS}$, are much less well constrained. Here, we lay out a program to evaluate this correction from effective field theory (EFT), highlighting the dominant terms as predicted by the EFT power counting. Moreover, we compare the results to a dispersive representation of $\delta_\text{NS}$ and show that the expected momentum scaling applies even in the case of low-lying intermediate states. Our EFT framework paves the way towards ab-initio calculations of $\delta_\text{NS}$ and thereby addresses the dominant uncertainty in $V_{ud}$.   
\end{abstract}

\maketitle

\emph{Introduction.}---A precise and robust determination of $V_{ud}$, the first element of the Cabibbo--Kobayashi--Maskawa (CKM) matrix~\cite{Cabibbo:1963yz,Kobayashi:1973fv}, is a critical input for the unitarity test of the first row of the CKM matrix
\begin{align}
\label{unitarity}
 |V_{ud}|^2+|V_{us}|^2+|V_{ub}|^2=1. 
\end{align}
At the moment, Eq.~\eqref{unitarity} displays a tension at the level of $2.8\sigma$~\cite{Cirigliano:2022yyo}. While a separate tension among $V_{us}$ determinations from kaon decays could potentially be resolved by future measurements at NA62~\cite{Cirigliano:2022yyo}, also $V_{ud}$ has come under increased scrutiny in recent years, mainly in view of the increased tension that followed from a reevaluation of universal radiative corrections (RC) associated with $\gamma W$ box diagrams~\cite{Marciano:2005ec,Seng:2018yzq,Seng:2018qru,Czarnecki:2019mwq,Seng:2020wjq,Hayen:2020cxh,Shiells:2020fqp}. Such a violation of CKM unitarity could point to a wide range of possible beyond-the-Standard-Model scenarios~\cite{Belfatto:2019swo,Coutinho:2019aiy}, including vector-like quarks~\cite{Cheung:2020vqm,Belfatto:2021jhf,Branco:2021vhs,Crivellin:2021bkd} and leptons~\cite{Crivellin:2020ebi,Kirk:2020wdk}, could be interpreted as a modification of the Fermi constant~\cite{Marciano:1999ih,Crivellin:2021njn}, the violation of lepton flavor universality~\cite{Crivellin:2020lzu,Crivellin:2020klg,Capdevila:2020rrl,Crivellin:2021sff,Crivellin:2020oup,Marzocca:2021azj}, or, more generally, in the context of Standard-Model EFT~\cite{Alok:2021ydy, Cirigliano:2022qdm,Cirigliano:2023nol,Dawid:2024wmp}. It is thus paramount to consolidate the evaluation of $V_{ud}$ and potentially even improve its precision. 

The current best determination  arises from superallowed $0^+\to 0^+$ transitions~\cite{Hardy:2020qwl}, for which the average over a large number of different isotopes ultimately yields the gain in precision compared to other probes. In those cases, the resulting precision of $V_{ud}$ is limited by experimental uncertainties: for neutron decay, recent years have witnessed impressive progress for the lifetime $\tau_n$~\cite{UCNt:2021pcg} and the decay parameter $\lambda$~\cite{Markisch:2018ndu}, but at least another factor of $2$ in the latter is required for a competitive determination, especially in view of the tension with Ref.~\cite{Beck:2019xye}. An extraction from pion $\beta$ decay would be theoretically even more pristine~\cite{Cirigliano:2002ng,Czarnecki:2019iwz,Feng:2020zdc}, yet experimentally challenging~\cite{Pocanic:2003pf}, forming a key physics goal of the PIONEER experiment~\cite{PIONEER:2022yag}.

In contrast, the challenges in the interpretation of superallowed $\beta$ decays are of theoretical nature. 
In the formula for the decay half-life $t$~\cite{Hardy:2020qwl,Gorchtein:2023naa}
\begin{equation}
    \frac{1}{t}  =    \frac{G_F^2 |V_{ud}|^2  m_e^5}{\pi^3 \log 2} (1 + \Delta^V_R)  (1 + \delta_R^\prime) (1 + \delta_\text{NS} - \delta_C)  \times f,  
    \label{eq:master0}
\end{equation}
 $f$ is a phase-space factor that includes the Fermi function, 
due to the Coulomb interaction of the outgoing electron  in the nuclear field, 
the nuclear electroweak (EW) form factor, nuclear recoil,  atomic electron screening, and  atomic overlap~\cite{Hardy:2020qwl,Gorchtein:2023naa}.
The other terms denote 
purely theoretical input
due to 
isospin-breaking and non-Coulomb RC. 
$\delta_C$ denotes the deviation of the  Fermi matrix element, 
$M_\text{F} = \langle f | \tau^+ |i \rangle = M_\text{F}^{(0)} (1 - \delta_C/2)$ from its isospin-limit value $M_\text{F}^{(0)} = \sqrt{2}$.
The so-called outer correction $\delta_R^\prime$ encodes all infrared-sensitive RC not included in the Fermi function.  At ${\mathcal O}(\alpha)$, these include the Sirlin function~\cite{Sirlin:1967zza}. The precise extraction of $V_{ud}$ requires control of corrections of ${\mathcal O}(\alpha^2 Z)$ and higher~\cite{Jaus:1972hua,Jaus:1986te,Sirlin:1986cc}.
The remaining RC are collectively denoted as the inner correction and  are  usually split into  
the single-nucleon correction $\Delta^V_R$  and the nuclear-structure-dependent term $\delta_\text{NS}$.
The latter arises from the phase-space average of a correction that in general depends on the positron energy $E_e$, as pointed 
out recently in Refs.~\cite{Gorchtein:2018fxl,Seng:2022cnq}. 

Currently,   the largest uncertainties reside in $\delta_C$ and $\delta_\text{NS}$. 
First, control over $\delta_C$ has long been a concern~\cite{Miller:2008my,Miller:2009cg} and Refs.~\cite{Martin:2021bud,Condren:2022dji,Seng:2022epj,Crawford:2022yhi,Seng:2023cvt,Seng:2023cgl}  
provide recent studies and strategies for improvements. 
Second, while the single-nucleon, universal RC from $\gamma W$ box diagrams have reached a good level of maturity, including a comprehensive analysis in EFT~\cite{Cirigliano:2023fnz} and a first lattice-QCD evaluation~\cite{Ma:2023kfr}, the same cannot be said for the  nucleus-dependent effects of the same diagrams. The nuclear correction called $\delta_\text{NS}$ dominates the resulting uncertainty in $V_{ud}$~\cite{Gorchtein:2018fxl}. The formalism for an evaluation using dispersion relations has been put forward in Refs.~\cite{Seng:2022cnq,Gorchtein:2023naa}, including subtleties that arise in the case of low-lying intermediate states, such as the $3^+$ and $1^+$ levels of $^{10}$B  in the $^{10}\text{C}\to{}^{10}\text{B}$ $0^+\to 0^+$ transition~\cite{Gennari:2024sbn}. 

In this Letter, we lay out a program to evaluate $\delta_\text{NS}$ in an EFT framework. We first set up the EFT power counting,  identify the leading contributions, 
and discuss the impact of low-lying nuclear states in the EFT and the
dispersive representation of Refs.~\cite{Seng:2022cnq,Gorchtein:2023naa}.
We then discuss in detail the leading nuclear-structure-dependent contribution $\delta_\text{NS}$. 
In particular, we analyze which contact terms are required as well 
as possible strategies for their determination, as has proved critical in the case of neutrinoless double-$\beta$ decay~\cite{Cirigliano:2018hja,Cirigliano:2019vdj,Cirigliano:2020dmx,Cirigliano:2021qko,Wirth:2021pij,Jokiniemi:2021qqv,Belley:2023lec}.

\emph{Effective field theory}.---The RC to nuclear $\beta$ decay involve several  widely separated energy scales. 
These range from the EW scale ($\mW$) to 
the very low-energy scale $q_\text{ext}$ of order of the reaction $\mathcal Q_\text{EC}$ value and  the electron mass $m_e$. 
The matrix element of the product of EW and electromagnetic (EM) currents 
in nuclear states brings in two additional  scales:  
the hadronic scale set by the nucleon mass $\mN$ (comparable to the breakdown scale of chiral perturbation theory, $\Lambda_\chi$) 
and the typical nuclear scales, $\gamma\simeq R^{-1} \simeq \mpi \simeq \mathcal O(100\MeV)$, with binding momentum $\gamma$, nuclear radius $R$, and pion mass $\mpi$.

In the spirit of EFT 
we exploit the hierarchy
\begin{equation}\label{eq:scales}
  q_\text{ext} \ll \mpi \ll  \Lambda_{\chi}  \ll \mW
\end{equation}
to systematically 
expand the $\beta$ decay amplitude in the ratios of 
scales probed by the virtual photon. Besides the ratio $G_F q^2_\text{ext}$ that sets the overall scale, these are  
\begin{equation}
    \epsilon_\text{recoil} = \mathcal O\bigg(\frac{q_\text{ext}}{\Lambda_\chi}\bigg), \quad  \epsilon_{\slashed{\pi}} = 
    \mathcal O\bigg(\frac{q_\text{ext}}{\mpi}\bigg), \quad
    \epsilon_{\chi} = 
    \mathcal O\bigg(\frac{\mpi}{\Lambda_\chi}\bigg),
 \end{equation}
 scaling roughly as $\simeq 
 0.005$, $\simeq 0.05$, and $\simeq 0.1$, respectively. 
Our goal is to catalog all corrections to superallowed $\beta$ decays at the permille level. 
This requires keeping $\mathcal O(\alpha \epsilon_\chi)$ and  $\mathcal O(\alpha \epsilon_{\slashed{\pi}})$ corrections, which are the focus of this Letter. Terms that are subleading in $\alpha$, but enhanced by the nuclear charge $Z$ or large logarithms, e.g., $\mathcal O(Z \alpha^2)$ or $\mathcal O(\alpha^2\log r)$, with $r$ a ratio of the scales in Eq.~\eqref{eq:scales}, are also relevant and discussed in detail in Ref.~\cite{Cirigliano:2024msg}, as is the potential role of $\Order(\alpha\epsilon_\chi^2)$ corrections, which are not yet included at present.

The presence of multiple scales requires the use of 
a tower of  EFTs, as done in 
the single nucleon sector~\cite{Cirigliano:2022hob,Cirigliano:2023fnz}. 
Between the EW scale and the hadronic scale, the relevant EFT is given by the Fermi theory 
obtained by integrating out the heavy Standard-Model particles.  
The resulting semileptonic operators are evolved 
using the  renormalization group (RG)
to the hadronic scale,  
where they are matched onto  an EFT  written in terms of  nucleons, pions, light leptons, and photons~\cite{Cirigliano:2022hob}, 
according to the symmetries of low-energy EW interactions, QED, and QCD. 

In terms of the heavy-baryon nucleon $N^T=(p,n)$  isodoublet,  the nucleon four-velocity $v_\mu$ and spin $S_\mu$,  and isospin Pauli matrices $\tau^a$~\cite{Jenkins:1990jv,Bernard:1995dp}, 
the leading-order (LO) EW one-body (1b) Lagrangian is 
\beq
   \mathcal L_W^\text{1b}   =
 - \sqrt{2} G_F V_{ud} \,   \bar e_L \gamma_\mu \nu_{L}
  \bar N   (g_V  v^\mu   - 2 g_A S^\mu ) \tau^+ N  + \cdots, 
\label{eq:Lagrangian_at_leading_order}
\eeq
where
the ellipsis denotes omitted terms involving pion fields or of  higher order in 
$\epsilon_\chi$. 
The effects of hard photons
with virtuality  $Q^2 \geq \Lambda_\chi^2$
 are captured in the deviation of the vector coupling $g_V$ from one (and $g_A$ from $g_A^\text{QCD}$~\cite{Cirigliano:2022hob}),
 see Ref.~\cite{Cirigliano:2024msg} for explicit expressions. 
Hard photons also generate EW two-body (2b) contact operators. 
We can write two $S$-wave operators relevant for superallowed $\beta$-decays that connect $^1S_0$ to $^1S_0$ states, with isospin $I=1$ and $I=2$, given by 
\begin{align}
\label{eq:2nuc1}
      \mathcal L_W^\text{2b}  & = -\sqrt{2} e^2 G_F V_{ud}  \bar e_L \slashed{v} \nu_L  \\  
    &\times    ( 
    g^{N\!N}_{V1}   N^\dagger \tau^+ N \, N^\dagger N  
     +  g^{N\!N}_{V2}  N^\dagger \tau^+ N \, N^\dagger \tau^3 N  
    ) + \cdots \notag
\end{align}
Weinberg power counting based on naive dimensional analysis would indicate that $g^{N\!N}_{V1,V2} = \mathcal O (\Lambda_\chi^{-3})$, but the requirement that the final nuclear amplitude be independent of the regulator  promotes the low-energy constants (LECs) to  $\mathcal O (\Lambda_\chi^{-1} \Fpi^{-2})$, where $\Fpi=92.3\MeV$ is the pion decay constant. The values of $g^{N\!N}_{V1,V2}$ are not known, but we will discuss strategies to obtain them below.

\begin{figure*}[ht]
 \includegraphics[width=0.95\textwidth]{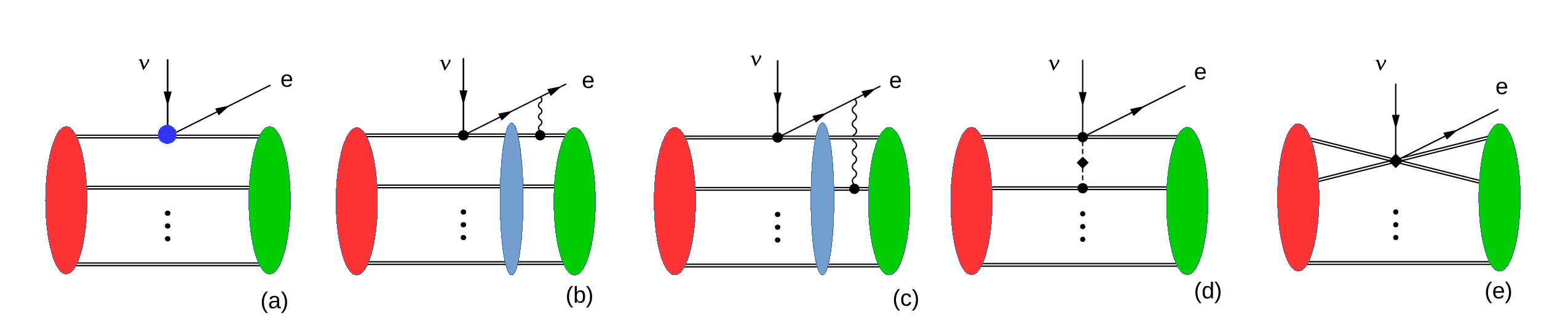}
\caption{Representative diagrams for RC to superallowed $\beta$ decays up to $\mathcal O(\alpha \eps_\chi)$ and $\mathcal O(\alpha \epsilon_{\slashed{\pi}})$. 
Leptons, nucleons, photons, and pions are denoted by plain, double, wavy, and dashed lines, respectively. A blue circle denotes the insertion of the EW current, including $\mathcal O(\alpha)$ corrections from hard photon exchange. Black circles denote 1b EW and EM currents. The red and green ovals denote the wave functions of the initial and final nuclei, the blue oval  represents the iteration of the nuclear interaction.}
\label{Fig:1}
\end{figure*}

Within this chiral EFT with dynamical photons and leptons we compute EW transition amplitudes 
involving multiple nucleons, see  Fig.~\ref{Fig:1} for some of the topologies relevant for nuclear decays.  
In the presence of more than one nucleon, the photon four-momentum can be in three  regions, see, e.g., Refs.~\cite{Beneke:1997zp,Smirnov:2002pj}:
\begin{enumerate}
    \item soft: $q^0_\gamma \simeq |\qg| \simeq \mpi$,
    \item ultrasoft: $q^0_\gamma \simeq |\qg| \simeq q_\text{ext}$,
    \item\label{potential} potential: $q^0_\gamma \simeq \qg^2/\mN \simeq q_\text{ext}$, $ |\qg| \simeq \mpi$. 
\end{enumerate}

In the nuclear  EFT, potential and soft modes are integrated out and give rise to an EW 
transition operator, analogous
to the pion-exchange potential in the strong sector.
In addition, hard photons also contribute to short-range transition operators proportional to $g^{N\!N}_{V1,V2}$ in Eq.~\eqref{eq:2nuc1}, see  diagram \ref{Fig:1}$(e)$.
Because the exchange of a hard, soft, or potential photon leaves the intermediate nuclear state far off-shell, these contributions can be calculated by taking the matrix element of the transition operator between the wave functions of the initial and final state. 
On the other hand, ultrasoft photons are sensitive to nuclear excitations and to the spectrum of intermediate states that are connected to the initial and final state by EW and EM currents. 

We now discuss the contributions from each region.
\\
\noindent {\it{Ultrasoft modes:}} 
Ultrasoft modes contribute at $\mathcal O(\alpha)$ through the LO photon--nucleon coupling.
Through topologies such as \ref{Fig:1}$(b)$ and \ref{Fig:1}$(c)$ 
(and real emission topologies that we have omitted), 
ultrasoft modes give rise to the Sirlin function~\cite{Sirlin:1967zza} 
and reproduce the $\mathcal O(\alpha)$ expansion of the
Fermi~\cite{Fermi:1934sk} function, with the correct nuclear charge $Z$, see Ref.~\cite{Cirigliano:2024msg} for details. Using known results, terms to all orders in $\alpha Z$, including logarithmically-enhanced terms that start at $\mathcal O(\alpha^2Z^2 \log \epsilon_{\slashed{\pi}})$, as well as terms at  $\mathcal O(\alpha^2 Z \log \epsilon_{\slashed{\pi}})$, and $\mathcal O(\alpha^2 \log \epsilon_{\slashed{\pi}})$ can be captured, see Refs.~\cite{Jaus:1970tah,Jaus:1972hua,Jaus:1986te,Sirlin:1977sv,Sirlin:1986cc,Sirlin:1986hpu}
and, in an EFT formalism, Refs.~\cite{Ando:2004rk,Cirigliano:2023fnz,Hill:2023acw,Hill:2023bfh,Borah:2024ghn}.
Subleading interactions, such as the interactions of the photon with the nucleon magnetic moment, are proportional to the ultrasoft momentum and appear at $\mathcal O(\alpha\epsilon_\text{recoil})$ beyond the order at which we work.\\ 
\noindent {\it Potential modes:}  Through topology \ref{Fig:1}$(c)$, 
potential modes give rise to $\mathcal O(\alpha \epsilon_{\slashed{\pi}})$
and $\mathcal O(\alpha \epsilon_{\chi})$ corrections to $\delta_\text{NS}$.  The former depend on the electron energy ($E_e$) and are induced by diagrams with the EW vector current and the EM charge density. The latter are $E_e$-independent and are induced by the axial current and the nucleon magnetic moments or recoil corrections to the vector current. Three-body (3b) potentials contribute at $\mathcal O(\alpha \epsilon_{\chi}^2)$ and are not shown.
\\
\noindent {\it Soft modes:}
Beyond tree level, the potentials 
receive corrections from one-loop diagrams involving soft pions and photons.  
By power counting, these 
first contribute to $\delta_\text{NS}$ at $\mathcal O(\alpha \epsilon^2_\chi )$
and $\mathcal O(\alpha^2)$. \\
\noindent {\it Hard modes:}
Hard modes give $\mathcal O(\alpha)$ corrections to $g_V$~\cite{Cirigliano:2023fnz} and generate  the $\mathcal O(\alpha \epsilon_\chi )$ 
two-nucleon counterterms 
 ($g^{N\!N}_{V1,V2}$)
needed for renormalization. In addition,  
they produce  $\mathcal O(\alpha \epsilon_{\slashed{\pi}}, \alpha \epsilon_\chi)$ effects in $\delta_\text{NS}$  through the electromagnetic
pion mass splitting in pion-mediated 2b currents, see diagram~\ref{Fig:1}$(d)$. The pion-mass splitting corrections are the nuclear analogs of the pion-induced RC in neutron decay~\cite{Cirigliano:2022hob}.

The  implication of this analysis is that 
in chiral EFT the dominant contribution to $\delta_\text{NS}$ comes from 
the matrix element of appropriate EW potentials between the initial and final nuclear states. Some contributions (from pion exchange) do not arise from  nuclear $\gamma W$ box diagrams. 
Sensitivity of $\delta_\text{NS}$ to intermediate nuclear states, including low-lying levels, arises from ultrasoft contributions 
that start to $\mathcal O (\alpha \epsilon_\text{recoil})$. This result is seemingly at odds 
with a recent dispersive analysis~\cite{Seng:2022cnq,Gorchtein:2023naa} 
in which some individual contributions scale as  $\mathcal O (\alpha \sqrt{\epsilon_\text{recoil}})$, and thus enhanced compared to the identified EFT scalings. We therefore turn next to a  detailed comparison to the dispersive representation.

\emph{Dispersive representation}.---In the current-algebra framework~\cite{Sirlin:1977sv} 
for EW RC, 
$\delta_\text{NS}$ arises from  the $\gamma W$ box diagram, in which a virtual photon is exchanged between 
the electron and the hadronic system. The relevant dynamical quantity is the Compton tensor
\begin{align}
\label{compton_tensor}
&T^{\mu\nu}(q;p',p)\notag\\
&=\frac{1}{2}\int \diff^4x\,e^{iq\cdot x}\langle f(p')|T\big\{J^\mu_\text{em}(x)J^\nu(0)\big\}|i(p)\rangle,
\end{align}
involving the matrix element of an EW and an EM current between the initial and final states with momentum $p$ and $p'$, respectively.
The $E_e$-independent part of $\delta_\text{NS}$ 
is induced by the  axial-vector component $T_A^{\mu\nu}$~\cite{Seng:2022cnq}.  
Ignoring recoil corrections, the relevant amplitude is expressed as the forward limit  
\beq
T_A^{\mu\nu}(p,q)=\frac{i\epsilon^{\mu\nu\alpha\beta}p_\alpha q_\beta}{2M\nu}T_3(\nu,Q^2),
\eeq
where $\nu=p\cdot q/M=q^0$, $Q^2=-q^2=-\nu^2+\qq^2$, and $M_i=M_f\equiv M$ has been assumed. 
Setting $m_e=0$,  the correction  relative to $M_\text{F}^{(0)}$
 becomes~\cite{Seng:2022cnq} 
\begin{align}
\label{box_gW}
\Box_{\gamma W}&=-\frac{e^2}{M_\text{F}^{(0)}}\int\frac{\diff^4q}{(2\pi)^4} \frac{\mW^2}{Q^2+\mW^2} 
\notag \\
&\times \frac{T_3(\nu,Q^2)}{(p_e-q)^2Q^2}\frac{Q^2+M\nu\frac{p_e\cdot q}{p\cdot p_e}}{M\nu},
\end{align}
where $p_e$ is the momentum of the positron. The nucleus-dependent correction is finally determined by subtracting 
the single-nucleon contribution, i.e.,~\cite{Seng:2022cnq} 
\beq
\label{deltaNS}
\delta_\text{NS}=2\big(\Box_{\gamma W}^\text{nucl}-\Box_{\gamma W}^n\big). 
\eeq

One way to perform the loop integral in Eq.~\eqref{box_gW}
relies on a Wick rotation $\nu\to i\nu_E$, which is advantageous when the Compton tensor is expressed via a dispersion relation.
However, as pointed out in Ref.~\cite{Gorchtein:2023naa}, such a Wick rotation is not always possible. In the case of low-lying nuclear states, as does happen in the $^{10}\text{C}\to{}^{10}\text{B}$ decay, the additional pole can move into the third quadrant and thus must be subtracted explicitly. It was found that the resulting residue contribution becomes singular for $E_e\to 0$, which could lead to a numerical enhancement. Such an enhancement should be reflected by the  momentum scaling and a different region in the EFT analysis.

To clarify the role of such low-lying states, 
we consider a simple example that displays all the relevant features
\beq
\frac{iT_3^\text{toy}(\nu,Q^2)}{M\nu}=\frac{M}{\mN}\frac{g_A g_M}{s-\bar M^2+i\eps},
\eeq
where $s=M^2+\nu^2-\qq^2+2M\nu$ and $M^2-\bar M^2=2M\Delta$. Here, $g_A$ and $g_M$ parameterize the matrix elements for the interaction with the EW and EM current, respectively. We focus on a single intermediate state with mass $\bar M$, with $\Delta>0$ corresponding to a low-lying state. The prefactor has been chosen to match the corresponding EFT expression~\cite{Cirigliano:2024msg}, counting the binding energy $\Delta\simeq q_\text{ext}$ as before.   

We can evaluate the integral by collecting all three residues in the upper half plane, see Ref.~\cite{Cirigliano:2024msg}, which gives
\beq
\label{toy_full}
\Box_{\gamma W}^{\text{toy}, \Delta}=\frac{3g_A g_M}{4M_\text{F}^{(0)}}\frac{\alpha}{\pi}\frac{\Delta}{\mN}\log\frac{2\Delta}{M}+\Order\big(\Delta^2\big),
\eeq
where we have only displayed the corrections to the $M\to \infty$ limit. As expected from the ultrasoft region in the EFT analysis, the result scales with $\mathcal O(\alpha\eps_\text{recoil})$.

In the dispersive approach,
the presence of a low-lying state impedes a straightforward Wick rotation, and its residue needs to be subtracted whenever the pole lies in the first or third quadrant. This gives rise to the residue contribution
\beq
\label{toy_res}
\Box_{\gamma W}^{\text{toy, res}}=\frac{g_A g_M}{M_\text{F}^{(0)}}\sqrt{\frac{M}{\mN}}\frac{\alpha}{\pi}\sqrt{\frac{2\Delta}{\mN}}+\Order\big(\Delta^{3/2}\big),
\eeq
which is again finite for $E_e\to 0$, but, contrary to Eq.~\eqref{toy_full}, scales as $\Order(\alpha\sqrt{\eps_\text{recoil}})$ and could thus be enhanced numerically. The solution to this apparent mismatch is that the Wick-rotated integral also involves terms scaling with $\sqrt{\Delta}$,
and one can show explicitly that~\cite{Cirigliano:2024msg}
\beq
\Box_{\gamma W}^\text{toy}=\Box_{\gamma W}^\text{toy, Wick}-\Box_{\gamma W}^\text{toy, res}.
\eeq
This demonstrates that no contributions of $\mathcal O(\alpha\sqrt{\eps_\text{recoil}})$ appear in the dispersive representation even in the case of low-lying states, confirming  the EFT scalings.

\emph{Leading contributions to $\delta_\emph{NS}$}.---In Ref.~\cite{Cirigliano:2024msg} we  derive  the nuclear decay rate in the EFT framework, 
while here we  focus on the implications for  $\delta_\text{NS}$.  
Potential modes induce an 
effective Hamiltonian of the form
\begin{equation}
\label{Heff}
    H_{\beta} = \sqrt{2} G_F V_{ud} \bar e_L \big[\gamma^0 (\mathcal V^0 + E_0 \mathcal V^0_{E} )  +  m_e \mathcal V_{m_e}  + \dots \big] \nu_L,
\end{equation}
where $E_0$ is the endpoint energy,
and the ellipsis denotes higher powers of lepton energy or $m_e$. The functions $\mathcal V^{0}$, 
$\mathcal V^{0}_E$, and $\mathcal V_{m_e}$ have a chiral expansion in $\epsilon_\chi$. The LO contributions to $H_\beta$ arise from diagrams such as those in Fig.~\ref{Fig:1}$(c,d,e)$.
We first consider  diagram~\ref{Fig:1}$(c)$. Because the LO 1b vector and axial currents are momentum independent, the LO potential is odd in the photon three-momentum $\qg$, and vanishes between $0^+$ states. To get a non-vanishing correction we need to retain the lepton momenta.
Similarly, the pion-exchange diagram~\ref{Fig:1}$(d)$ involving LO vertices requires an insertion of an external lepton momentum, leading to the only non-vanishing LO contributions
\begin{align}\label{eq:FermiPot}
    \mathcal V^0_E &= \frac{1}{3}\left( \frac{1}{2} +  \frac{4 E_e}{E_0}\right) \mathcal V_E + \mathcal V^{\pi}_{E}, \quad
    \mathcal V_{m_e}  = \frac{1}{2} \mathcal V_E + \mathcal V^{\pi}_{m_e},
\end{align}
with the explicit expressions for $\mathcal V_E$, $\mathcal V_{E}^\pi$, and $\mathcal V_{m_e}^\pi$ given in Ref.~\cite{Cirigliano:2024msg}. The latter two depend on the pion mass splitting, $M_{\pi^\pm}^2 - M_{\pi^0}^2 = 2 e^2  F_\pi^2 Z_\pi$, encoding effects of hard photons. 
 These potentials are energy dependent and affect both the spectral shape and the total decay rate at $\mathcal O(\alpha \eps_{\slashed{\pi}})$. 

Additional contributions arise from diagrams~\ref{Fig:1}$(c,d)$ when using subleading vertices instead of inserting a lepton momentum. One order down in the chiral expansion at $\mathcal O(\alpha \eps_\chi)$ we obtain potentials that are independent of the lepton momenta
and thus contribute to $\mathcal V^0$. These
$\mathcal O(\alpha \epsilon_\chi)$
terms can be further decomposed as
\begin{equation}
\label{eq:MagPotential}
    \mathcal V^0 = \mathcal V^\text{mag}_0  + \mathcal V^\text{rec}_0 + \mathcal V^\text{CT}_0, 
\end{equation}
corresponding to diagrams~\ref{Fig:1}$(c,d)$ via magnetic, recoil, and contact-term contributions~\cite{Cirigliano:2024msg}.
Beyond tree level, the potentials~\eqref{eq:FermiPot}
and~\eqref{eq:MagPotential}
receive corrections from soft pions and photons at $\mathcal O(\alpha \epsilon_\chi^2)$ and $\mathcal O(\alpha^2)$, beyond the accuracy of this work. At this order we also expect effects from 3b potentials.

It is important to notice that $\mathcal V^\text{mag}_0$ has a Coulombic, $\qq^{-2} $, scaling. Such a potential, when inserted into ${}^1S_0$ chiral EFT wave functions, gives rise to nuclear matrix elements that depend logarithmically on the applied regulator~\cite{PavonValderrama:2014zeq,Cirigliano:2018hja}. This regulator dependence signals sensitivity to hard-photon exchange between nucleons which, in chiral EFT, are captured by the short-range operators in Eq.~\eqref{eq:2nuc1}. The corresponding LECs absorb the regulator dependence and after renormalization are enhanced over naive dimensional analysis as anticipated below Eq.~\eqref{eq:2nuc1}. This is analogous to the short-range operators identified for neutrinoless double-$\beta$ decay~\cite{Cirigliano:2018hja,Cirigliano:2019vdj}. The short-range terms give an $\mathcal O(\alpha \eps_\chi)$ contribution
\beq
\mathcal V_0^\text{CT} = e^2 \big(g^{N\!N}_{V1} O_1 + g^{N\!N}_{V2} O_2\big),
\eeq
where 
\beq
O_1 = \sum_{j\neq k} \tau^{+(j)}\mathds{1}_k,\quad O_2 = \sum_{j<k} \big[\tau^{+(j)}\tau_3^{(k)}+(j\leftrightarrow k)\big].
\eeq
$\mathcal V_0^\text{CT}$ depends on two unknown LECs and corresponds to genuine new 2b contributions arising from high-momentum photon exchange. It is an intrinsic two-nucleon effect that cannot be obtained from one-nucleon processes. Below we compute the contributions of $g^{N\!N}_{V1,V2}$ by using the scaling discussed below Eq.~\eqref{eq:2nuc1} for the LECs and treating the result as an overall uncertainty \cite{Cirigliano:2024msg}.

In the EFT approach,  up to the order considered, $\delta_\text{NS}$ is  entirely determined by matrix elements 
of appropriate potentials, see Eq.~\eqref{Heff},  between the initial and final states without dependence on intermediate nuclear states.
The EFT power counting indicates that $\delta_\text{NS}$ 
receives a LO $E_e$-independent contribution of  $\mathcal O(\alpha \epsilon_\chi)$, $\delta_\text{NS}^{(0)}$, 
and an $E_e$-dependent contribution of $\mathcal O(\alpha \epsilon_{\slashed{\pi}})$, $\delta^E_\text{NS}$.  
In the case of $\delta_\text{NS}^{(0)}$ we also found an $\mathcal O(\alpha^2)$ potential $\mathcal V_+$ that needs to be included for $\mathcal O(10^{-4})$ precision~\cite{Cirigliano:2024msg}. 

The  two currently unknown LECs $g^{N\!N}_{V1,V2}$ can be determined in the future  both from theory and experiment. 
First, one can envision  a matching calculation to the underlying theory, 
performed in lattice QCD or  within the Cottingham-like approach~\cite{Cirigliano:2020dmx,Cirigliano:2021qko}.  
Second, the LECs can be extracted from experimental data, 
based on the observations  that: (i) there are ${\mathcal O}(10)$ very precisely measured superallowed $\beta$ decays \cite{Hardy:2020qwl}, 
connecting members of $I=1$ triplets with initial $m_I=-1$ or $m_I=0$; 
(ii) the LECs  contribute to $\delta_\text{NS}$ through  the combinations  $g^{N\!N}_{V1} \langle f||O_1||i\rangle \mp \sqrt{3/5} g^{N\!N}_{V2} \langle f||O_2||i\rangle$,  
depending  on whether  $m_I=-1$ or $m_I=0$, and $\langle f||O_{1,2}||i\rangle$ are reduced matrix elements that depend on the decaying nucleus and can be computed with ab-initio nuclear methods.  It is then possible to perform a global fit to extract values of $g^{N\!N}_{V1,V2}$ and $V_{ud}$ simultaneously from the set of superallowed $\beta$ decay measurements.

Based on the EFT framework described here, we  have derived 
a master formula for the decay rate  and performed  
 first  numerical calculations for $\delta_\text{NS}$ in the decay 
 $^{14}\text{O}\to{}^{14}\text{N}$  with  Quantum Monte Carlo methods, 
 confirming the expectations from the EFT power counting~\cite{Cirigliano:2024msg}. 
As an illustration, we extract  $V_{ud}$ from the $^{14}\text{O}$ decay,  finding 
 $V_{ud} \big[{}^{14}\text{O}\big]
 =0.97364(56)$, 
 with uncertainty dominated by our ignorance of the LECs, $(\delta V_{ud} )_{g_V^{N\!N}} = 4.3 \times 10^{-4}$. 
 Eliminating this uncertainty 
  would result in $\delta V_{ud} = 3.6 \times 10^{-4}$. 
This is to be compared with $V_{ud} \big[{}^{14}\text{O}\big]= 0.97405(37)$ from Ref.~\cite{Hardy:2020qwl}, 
 with uncertainty dominated by $\delta_\text{NS}$,   $(\delta V_{ud} )_{\delta_\text{NS}} = 3.1 \times 10^{-4}$, 
 and with  $V_{ud} =  0.97373(31)$ obtained by a global analysis of the $0^+ \to 0^+$ decays~\cite{Hardy:2020qwl}.
These considerations show that there is a clear path towards reaching $\delta V_{ud} \simeq 3 \times 10^{-4}$,  
once the LECs are determined following the strategies outlined above.  
We expect that a few  decays of  light nuclei, combined with nuclear-structure calculations, should suffice to obtain
 a competitive determination of $V_{ud}$, including a robust estimate of the nuclear-structure uncertainties.

\emph{Discussion and outlook}.---We have performed a first study  of RC to superallowed nuclear $\beta$ decays in 
an  EFT framework that bridges the EW scale to nuclear scales. 
We have identified the leading nuclear-structure-dependent corrections $\delta_\text{NS}$ 
as arising from matrix elements of EW transition operators of 
$\mathcal O(G_F \alpha \epsilon_{\slashed{\pi}}, G_F \alpha \epsilon_\chi)$ 
between initial and final nuclear wave functions.
Several terms, such as the magnetic and recoil pieces of $\delta_\text{NS}$, already appear in the seminal work~\cite{Towner:1992xm}, while others are new. Most strikingly, we identified novel pion-exchange and short-range corrections that affect $\delta_\text{NS}$ at the same order as the usually considered corrections. Furthermore, we have sketched a strategy
using global fits to superallowed $\beta$ decays to empirically determine
the contact operators' Wilson coefficients. 

To map these EFT considerations onto a dispersive approach for $\delta_\text{NS}$~\cite{Gorchtein:2018fxl,Seng:2022cnq}, we first showed that the only contributions that scale with $q_\text{ext}$ arise in the potential region, and thus do not depend on the properties of individual states. This remains true in the presence of low-lying levels. 
Second,  while the leading $\mathcal O(\alpha \epsilon_\chi)$ effects are energy independent, $\mathcal O(\alpha \epsilon_{\slashed{\pi}})$  energy-dependent corrections are predicted by the EFT, related to $\delta_\text{NS}^E$ in the dispersive approach.

In conclusion, the EFT approach  presented in this Letter allows one to derive corrections in a systematic way, and thereby opens up new avenues 
 to control the theoretical uncertainties in superallowed nuclear $\beta$ decays. This  enables first-principles nuclear many-body calculations of structure-dependent corrections, whose uncertainty 
currently dominates the extraction of $V_{ud}$, to further sharpen precision tests of the Standard Model and potentially reveal hints of physics beyond.

\begin{acknowledgments}
We thank Mikhail Gorchtein, Vaisakh Plakkot,  Chien-Yeah Seng,
and Oleksandr Tomalak  for valuable discussions. Financial support by the Dutch Research Council (NWO) in the form of a VIDI grant, the U.S.\ DOE (Grant No.\
DE-FG02-00ER41132), Los Alamos
National Laboratory's Laboratory Directed Research and Development program under projects
20210190ER and 20210041DR, and the  SNSF (Project No.\ PCEFP2\_181117) is gratefully acknowledged.  Los Alamos National Laboratory is operated by Triad National Security, LLC,
for the National Nuclear Security Administration of U.S.\ Department of Energy (Contract No.\
89233218CNA000001). 
We acknowledge support from the DOE Topical Collaboration ``Nuclear Theory for New Physics,'' award No.\ DE-SC0023663. The work of S.G.\ is also supported by the Office of Advanced Scientific Computing Research, Scientific Discovery through Advanced Computing (SciDAC) NUCLEI program
and  by the Network for Neutrinos, Nuclear Astrophysics, and Symmetries (N3AS).
This research used resources provided by the Los Alamos National Laboratory Institutional Computing Program, which is supported by the U.S.\ Department of Energy National Nuclear Security Administration under Contract No.\ 89233218CNA000001. M.H.\ and E.M.\ thank the Institute for Nuclear Theory at the University of Washington for its kind hospitality and support during the program ``New physics searches at the precision frontier (INT-23-1b),'' when this project was initiated. 
\end{acknowledgments}

\bibliography{residue}

\begin{thebibliography}{80}
\expandafter\ifx\csname natexlab\endcsname\relax\def\natexlab#1{#1}\fi
\expandafter\ifx\csname bibnamefont\endcsname\relax
  \def\bibnamefont#1{#1}\fi
\expandafter\ifx\csname bibfnamefont\endcsname\relax
  \def\bibfnamefont#1{#1}\fi
\expandafter\ifx\csname citenamefont\endcsname\relax
  \def\citenamefont#1{#1}\fi
\expandafter\ifx\csname url\endcsname\relax
  \def\url#1{\texttt{#1}}\fi
\expandafter\ifx\csname urlprefix\endcsname\relax\def\urlprefix{URL }\fi
\providecommand{\bibinfo}[2]{#2}
\providecommand{\eprint}[2][]{\url{#2}}

\bibitem[{\citenamefont{Cabibbo}(1963)}]{Cabibbo:1963yz}
\bibinfo{author}{\bibfnamefont{N.}~\bibnamefont{Cabibbo}},
  \bibinfo{journal}{Phys. Rev. Lett.} \textbf{\bibinfo{volume}{10}},
  \bibinfo{pages}{531} (\bibinfo{year}{1963}).

\bibitem[{\citenamefont{Kobayashi and Maskawa}(1973)}]{Kobayashi:1973fv}
\bibinfo{author}{\bibfnamefont{M.}~\bibnamefont{Kobayashi}} \bibnamefont{and}
  \bibinfo{author}{\bibfnamefont{T.}~\bibnamefont{Maskawa}},
  \bibinfo{journal}{Prog. Theor. Phys.} \textbf{\bibinfo{volume}{49}},
  \bibinfo{pages}{652} (\bibinfo{year}{1973}).

\bibitem[{\citenamefont{Cirigliano
  et~al.}(2023{\natexlab{a}})\citenamefont{Cirigliano, Crivellin, Hoferichter,
  and Moulson}}]{Cirigliano:2022yyo}
\bibinfo{author}{\bibfnamefont{V.}~\bibnamefont{Cirigliano}},
  \bibinfo{author}{\bibfnamefont{A.}~\bibnamefont{Crivellin}},
  \bibinfo{author}{\bibfnamefont{M.}~\bibnamefont{Hoferichter}},
  \bibnamefont{and} \bibinfo{author}{\bibfnamefont{M.}~\bibnamefont{Moulson}},
  \bibinfo{journal}{Phys. Lett. B} \textbf{\bibinfo{volume}{838}},
  \bibinfo{pages}{137748} (\bibinfo{year}{2023}{\natexlab{a}}),
  \eprint{2208.11707}.

\bibitem[{\citenamefont{Marciano and Sirlin}(2006)}]{Marciano:2005ec}
\bibinfo{author}{\bibfnamefont{W.~J.} \bibnamefont{Marciano}} \bibnamefont{and}
  \bibinfo{author}{\bibfnamefont{A.}~\bibnamefont{Sirlin}},
  \bibinfo{journal}{Phys. Rev. Lett.} \textbf{\bibinfo{volume}{96}},
  \bibinfo{pages}{032002} (\bibinfo{year}{2006}), \eprint{hep-ph/0510099}.

\bibitem[{\citenamefont{Seng et~al.}(2018)\citenamefont{Seng, Gorchtein, Patel,
  and Ramsey-Musolf}}]{Seng:2018yzq}
\bibinfo{author}{\bibfnamefont{C.-Y.} \bibnamefont{Seng}},
  \bibinfo{author}{\bibfnamefont{M.}~\bibnamefont{Gorchtein}},
  \bibinfo{author}{\bibfnamefont{H.~H.} \bibnamefont{Patel}}, \bibnamefont{and}
  \bibinfo{author}{\bibfnamefont{M.~J.} \bibnamefont{Ramsey-Musolf}},
  \bibinfo{journal}{Phys. Rev. Lett.} \textbf{\bibinfo{volume}{121}},
  \bibinfo{pages}{241804} (\bibinfo{year}{2018}), \eprint{1807.10197}.

\bibitem[{\citenamefont{Seng et~al.}(2019)\citenamefont{Seng, Gorchtein, and
  Ramsey-Musolf}}]{Seng:2018qru}
\bibinfo{author}{\bibfnamefont{C.~Y.} \bibnamefont{Seng}},
  \bibinfo{author}{\bibfnamefont{M.}~\bibnamefont{Gorchtein}},
  \bibnamefont{and} \bibinfo{author}{\bibfnamefont{M.~J.}
  \bibnamefont{Ramsey-Musolf}}, \bibinfo{journal}{Phys. Rev. D}
  \textbf{\bibinfo{volume}{100}}, \bibinfo{pages}{013001}
  (\bibinfo{year}{2019}), \eprint{1812.03352}.

\bibitem[{\citenamefont{Czarnecki et~al.}(2019)\citenamefont{Czarnecki,
  Marciano, and Sirlin}}]{Czarnecki:2019mwq}
\bibinfo{author}{\bibfnamefont{A.}~\bibnamefont{Czarnecki}},
  \bibinfo{author}{\bibfnamefont{W.~J.} \bibnamefont{Marciano}},
  \bibnamefont{and} \bibinfo{author}{\bibfnamefont{A.}~\bibnamefont{Sirlin}},
  \bibinfo{journal}{Phys. Rev. D} \textbf{\bibinfo{volume}{100}},
  \bibinfo{pages}{073008} (\bibinfo{year}{2019}), \eprint{1907.06737}.

\bibitem[{\citenamefont{Seng et~al.}(2020)\citenamefont{Seng, Feng, Gorchtein,
  and Jin}}]{Seng:2020wjq}
\bibinfo{author}{\bibfnamefont{C.-Y.} \bibnamefont{Seng}},
  \bibinfo{author}{\bibfnamefont{X.}~\bibnamefont{Feng}},
  \bibinfo{author}{\bibfnamefont{M.}~\bibnamefont{Gorchtein}},
  \bibnamefont{and} \bibinfo{author}{\bibfnamefont{L.-C.} \bibnamefont{Jin}},
  \bibinfo{journal}{Phys. Rev. D} \textbf{\bibinfo{volume}{101}},
  \bibinfo{pages}{111301} (\bibinfo{year}{2020}), \eprint{2003.11264}.

\bibitem[{\citenamefont{Hayen}(2021)}]{Hayen:2020cxh}
\bibinfo{author}{\bibfnamefont{L.}~\bibnamefont{Hayen}},
  \bibinfo{journal}{Phys. Rev. D} \textbf{\bibinfo{volume}{103}},
  \bibinfo{pages}{113001} (\bibinfo{year}{2021}), \eprint{2010.07262}.

\bibitem[{\citenamefont{Shiells et~al.}(2021)\citenamefont{Shiells, Blunden,
  and Melnitchouk}}]{Shiells:2020fqp}
\bibinfo{author}{\bibfnamefont{K.}~\bibnamefont{Shiells}},
  \bibinfo{author}{\bibfnamefont{P.~G.} \bibnamefont{Blunden}},
  \bibnamefont{and}
  \bibinfo{author}{\bibfnamefont{W.}~\bibnamefont{Melnitchouk}},
  \bibinfo{journal}{Phys. Rev. D} \textbf{\bibinfo{volume}{104}},
  \bibinfo{pages}{033003} (\bibinfo{year}{2021}), \eprint{2012.01580}.

\bibitem[{\citenamefont{Belfatto et~al.}(2020)\citenamefont{Belfatto, Beradze,
  and Berezhiani}}]{Belfatto:2019swo}
\bibinfo{author}{\bibfnamefont{B.}~\bibnamefont{Belfatto}},
  \bibinfo{author}{\bibfnamefont{R.}~\bibnamefont{Beradze}}, \bibnamefont{and}
  \bibinfo{author}{\bibfnamefont{Z.}~\bibnamefont{Berezhiani}},
  \bibinfo{journal}{Eur. Phys. J. C} \textbf{\bibinfo{volume}{80}},
  \bibinfo{pages}{149} (\bibinfo{year}{2020}), \eprint{1906.02714}.

\bibitem[{\citenamefont{Coutinho et~al.}(2020)\citenamefont{Coutinho,
  Crivellin, and Manzari}}]{Coutinho:2019aiy}
\bibinfo{author}{\bibfnamefont{A.~M.} \bibnamefont{Coutinho}},
  \bibinfo{author}{\bibfnamefont{A.}~\bibnamefont{Crivellin}},
  \bibnamefont{and} \bibinfo{author}{\bibfnamefont{C.~A.}
  \bibnamefont{Manzari}}, \bibinfo{journal}{Phys. Rev. Lett.}
  \textbf{\bibinfo{volume}{125}}, \bibinfo{pages}{071802}
  (\bibinfo{year}{2020}), \eprint{1912.08823}.

\bibitem[{\citenamefont{Cheung et~al.}(2020)\citenamefont{Cheung, Keung, Lu,
  and Tseng}}]{Cheung:2020vqm}
\bibinfo{author}{\bibfnamefont{K.}~\bibnamefont{Cheung}},
  \bibinfo{author}{\bibfnamefont{W.-Y.} \bibnamefont{Keung}},
  \bibinfo{author}{\bibfnamefont{C.-T.} \bibnamefont{Lu}}, \bibnamefont{and}
  \bibinfo{author}{\bibfnamefont{P.-Y.} \bibnamefont{Tseng}},
  \bibinfo{journal}{JHEP} \textbf{\bibinfo{volume}{05}}, \bibinfo{pages}{117}
  (\bibinfo{year}{2020}), \eprint{2001.02853}.

\bibitem[{\citenamefont{Belfatto and Berezhiani}(2021)}]{Belfatto:2021jhf}
\bibinfo{author}{\bibfnamefont{B.}~\bibnamefont{Belfatto}} \bibnamefont{and}
  \bibinfo{author}{\bibfnamefont{Z.}~\bibnamefont{Berezhiani}},
  \bibinfo{journal}{JHEP} \textbf{\bibinfo{volume}{10}}, \bibinfo{pages}{079}
  (\bibinfo{year}{2021}), \eprint{2103.05549}.

\bibitem[{\citenamefont{Branco et~al.}(2021)\citenamefont{Branco, Penedo,
  Pereira, Rebelo, and Silva-Marcos}}]{Branco:2021vhs}
\bibinfo{author}{\bibfnamefont{G.~C.} \bibnamefont{Branco}},
  \bibinfo{author}{\bibfnamefont{J.~T.} \bibnamefont{Penedo}},
  \bibinfo{author}{\bibfnamefont{P.~M.~F.} \bibnamefont{Pereira}},
  \bibinfo{author}{\bibfnamefont{M.~N.} \bibnamefont{Rebelo}},
  \bibnamefont{and} \bibinfo{author}{\bibfnamefont{J.~I.}
  \bibnamefont{Silva-Marcos}}, \bibinfo{journal}{JHEP}
  \textbf{\bibinfo{volume}{07}}, \bibinfo{pages}{099} (\bibinfo{year}{2021}),
  \eprint{2103.13409}.

\bibitem[{\citenamefont{Crivellin
  et~al.}(2021{\natexlab{a}})\citenamefont{Crivellin, Hoferichter, Kirk,
  Manzari, and Schnell}}]{Crivellin:2021bkd}
\bibinfo{author}{\bibfnamefont{A.}~\bibnamefont{Crivellin}},
  \bibinfo{author}{\bibfnamefont{M.}~\bibnamefont{Hoferichter}},
  \bibinfo{author}{\bibfnamefont{M.}~\bibnamefont{Kirk}},
  \bibinfo{author}{\bibfnamefont{C.~A.} \bibnamefont{Manzari}},
  \bibnamefont{and} \bibinfo{author}{\bibfnamefont{L.}~\bibnamefont{Schnell}},
  \bibinfo{journal}{JHEP} \textbf{\bibinfo{volume}{10}}, \bibinfo{pages}{221}
  (\bibinfo{year}{2021}{\natexlab{a}}), \eprint{2107.13569}.

\bibitem[{\citenamefont{Crivellin et~al.}(2020)\citenamefont{Crivellin, Kirk,
  Manzari, and Montull}}]{Crivellin:2020ebi}
\bibinfo{author}{\bibfnamefont{A.}~\bibnamefont{Crivellin}},
  \bibinfo{author}{\bibfnamefont{F.}~\bibnamefont{Kirk}},
  \bibinfo{author}{\bibfnamefont{C.~A.} \bibnamefont{Manzari}},
  \bibnamefont{and} \bibinfo{author}{\bibfnamefont{M.}~\bibnamefont{Montull}},
  \bibinfo{journal}{JHEP} \textbf{\bibinfo{volume}{12}}, \bibinfo{pages}{166}
  (\bibinfo{year}{2020}), \eprint{2008.01113}.

\bibitem[{\citenamefont{Kirk}(2021)}]{Kirk:2020wdk}
\bibinfo{author}{\bibfnamefont{M.}~\bibnamefont{Kirk}}, \bibinfo{journal}{Phys.
  Rev. D} \textbf{\bibinfo{volume}{103}}, \bibinfo{pages}{035004}
  (\bibinfo{year}{2021}), \eprint{2008.03261}.

\bibitem[{\citenamefont{Marciano}(1999)}]{Marciano:1999ih}
\bibinfo{author}{\bibfnamefont{W.~J.} \bibnamefont{Marciano}},
  \bibinfo{journal}{Phys. Rev. D} \textbf{\bibinfo{volume}{60}},
  \bibinfo{pages}{093006} (\bibinfo{year}{1999}), \eprint{hep-ph/9903451}.

\bibitem[{\citenamefont{Crivellin
  et~al.}(2021{\natexlab{b}})\citenamefont{Crivellin, Hoferichter, and
  Manzari}}]{Crivellin:2021njn}
\bibinfo{author}{\bibfnamefont{A.}~\bibnamefont{Crivellin}},
  \bibinfo{author}{\bibfnamefont{M.}~\bibnamefont{Hoferichter}},
  \bibnamefont{and} \bibinfo{author}{\bibfnamefont{C.~A.}
  \bibnamefont{Manzari}}, \bibinfo{journal}{Phys. Rev. Lett.}
  \textbf{\bibinfo{volume}{127}}, \bibinfo{pages}{071801}
  (\bibinfo{year}{2021}{\natexlab{b}}), \eprint{2102.02825}.

\bibitem[{\citenamefont{Crivellin and Hoferichter}(2020)}]{Crivellin:2020lzu}
\bibinfo{author}{\bibfnamefont{A.}~\bibnamefont{Crivellin}} \bibnamefont{and}
  \bibinfo{author}{\bibfnamefont{M.}~\bibnamefont{Hoferichter}},
  \bibinfo{journal}{Phys. Rev. Lett.} \textbf{\bibinfo{volume}{125}},
  \bibinfo{pages}{111801} (\bibinfo{year}{2020}), \eprint{2002.07184}.

\bibitem[{\citenamefont{Crivellin
  et~al.}(2021{\natexlab{c}})\citenamefont{Crivellin, Kirk, Manzari, and
  Panizzi}}]{Crivellin:2020klg}
\bibinfo{author}{\bibfnamefont{A.}~\bibnamefont{Crivellin}},
  \bibinfo{author}{\bibfnamefont{F.}~\bibnamefont{Kirk}},
  \bibinfo{author}{\bibfnamefont{C.~A.} \bibnamefont{Manzari}},
  \bibnamefont{and} \bibinfo{author}{\bibfnamefont{L.}~\bibnamefont{Panizzi}},
  \bibinfo{journal}{Phys. Rev. D} \textbf{\bibinfo{volume}{103}},
  \bibinfo{pages}{073002} (\bibinfo{year}{2021}{\natexlab{c}}),
  \eprint{2012.09845}.

\bibitem[{\citenamefont{Capdevila et~al.}(2021)\citenamefont{Capdevila,
  Crivellin, Manzari, and Montull}}]{Capdevila:2020rrl}
\bibinfo{author}{\bibfnamefont{B.}~\bibnamefont{Capdevila}},
  \bibinfo{author}{\bibfnamefont{A.}~\bibnamefont{Crivellin}},
  \bibinfo{author}{\bibfnamefont{C.~A.} \bibnamefont{Manzari}},
  \bibnamefont{and} \bibinfo{author}{\bibfnamefont{M.}~\bibnamefont{Montull}},
  \bibinfo{journal}{Phys. Rev. D} \textbf{\bibinfo{volume}{103}},
  \bibinfo{pages}{015032} (\bibinfo{year}{2021}), \eprint{2005.13542}.

\bibitem[{\citenamefont{Crivellin and Hoferichter}(2021)}]{Crivellin:2021sff}
\bibinfo{author}{\bibfnamefont{A.}~\bibnamefont{Crivellin}} \bibnamefont{and}
  \bibinfo{author}{\bibfnamefont{M.}~\bibnamefont{Hoferichter}},
  \bibinfo{journal}{Science} \textbf{\bibinfo{volume}{374}},
  \bibinfo{pages}{1051} (\bibinfo{year}{2021}), \eprint{2111.12739}.

\bibitem[{\citenamefont{Crivellin
  et~al.}(2021{\natexlab{d}})\citenamefont{Crivellin, Manzari, Alguer{\'o}, and
  Matias}}]{Crivellin:2020oup}
\bibinfo{author}{\bibfnamefont{A.}~\bibnamefont{Crivellin}},
  \bibinfo{author}{\bibfnamefont{C.~A.} \bibnamefont{Manzari}},
  \bibinfo{author}{\bibfnamefont{M.}~\bibnamefont{Alguer{\'o}}},
  \bibnamefont{and} \bibinfo{author}{\bibfnamefont{J.}~\bibnamefont{Matias}},
  \bibinfo{journal}{Phys. Rev. Lett.} \textbf{\bibinfo{volume}{127}},
  \bibinfo{pages}{011801} (\bibinfo{year}{2021}{\natexlab{d}}),
  \eprint{2010.14504}.

\bibitem[{\citenamefont{Marzocca and Trifinopoulos}(2021)}]{Marzocca:2021azj}
\bibinfo{author}{\bibfnamefont{D.}~\bibnamefont{Marzocca}} \bibnamefont{and}
  \bibinfo{author}{\bibfnamefont{S.}~\bibnamefont{Trifinopoulos}},
  \bibinfo{journal}{Phys. Rev. Lett.} \textbf{\bibinfo{volume}{127}},
  \bibinfo{pages}{061803} (\bibinfo{year}{2021}), \eprint{2104.05730}.

\bibitem[{\citenamefont{Alok et~al.}(2023)\citenamefont{Alok, Dighe, Gangal,
  and Kumar}}]{Alok:2021ydy}
\bibinfo{author}{\bibfnamefont{A.~K.} \bibnamefont{Alok}},
  \bibinfo{author}{\bibfnamefont{A.}~\bibnamefont{Dighe}},
  \bibinfo{author}{\bibfnamefont{S.}~\bibnamefont{Gangal}}, \bibnamefont{and}
  \bibinfo{author}{\bibfnamefont{J.}~\bibnamefont{Kumar}},
  \bibinfo{journal}{Phys. Rev. D} \textbf{\bibinfo{volume}{108}},
  \bibinfo{pages}{113005} (\bibinfo{year}{2023}), \eprint{2108.05614}.

\bibitem[{\citenamefont{Cirigliano
  et~al.}(2022{\natexlab{a}})\citenamefont{Cirigliano, Dekens, de~Vries,
  Mereghetti, and Tong}}]{Cirigliano:2022qdm}
\bibinfo{author}{\bibfnamefont{V.}~\bibnamefont{Cirigliano}},
  \bibinfo{author}{\bibfnamefont{W.}~\bibnamefont{Dekens}},
  \bibinfo{author}{\bibfnamefont{J.}~\bibnamefont{de~Vries}},
  \bibinfo{author}{\bibfnamefont{E.}~\bibnamefont{Mereghetti}},
  \bibnamefont{and} \bibinfo{author}{\bibfnamefont{T.}~\bibnamefont{Tong}},
  \bibinfo{journal}{Phys. Rev. D} \textbf{\bibinfo{volume}{106}},
  \bibinfo{pages}{075001} (\bibinfo{year}{2022}{\natexlab{a}}),
  \eprint{2204.08440}.

\bibitem[{\citenamefont{Cirigliano
  et~al.}(2024{\natexlab{a}})\citenamefont{Cirigliano, Dekens, de~Vries,
  Mereghetti, and Tong}}]{Cirigliano:2023nol}
\bibinfo{author}{\bibfnamefont{V.}~\bibnamefont{Cirigliano}},
  \bibinfo{author}{\bibfnamefont{W.}~\bibnamefont{Dekens}},
  \bibinfo{author}{\bibfnamefont{J.}~\bibnamefont{de~Vries}},
  \bibinfo{author}{\bibfnamefont{E.}~\bibnamefont{Mereghetti}},
  \bibnamefont{and} \bibinfo{author}{\bibfnamefont{T.}~\bibnamefont{Tong}},
  \bibinfo{journal}{JHEP} \textbf{\bibinfo{volume}{03}}, \bibinfo{pages}{033}
  (\bibinfo{year}{2024}{\natexlab{a}}), \eprint{2311.00021}.

\bibitem[{\citenamefont{Dawid et~al.}(2024)\citenamefont{Dawid, Cirigliano, and
  Dekens}}]{Dawid:2024wmp}
\bibinfo{author}{\bibfnamefont{M.}~\bibnamefont{Dawid}},
  \bibinfo{author}{\bibfnamefont{V.}~\bibnamefont{Cirigliano}},
  \bibnamefont{and} \bibinfo{author}{\bibfnamefont{W.}~\bibnamefont{Dekens}},
  \bibinfo{journal}{JHEP} \textbf{\bibinfo{volume}{08}}, \bibinfo{pages}{175}
  (\bibinfo{year}{2024}), \eprint{2402.06723}.

\bibitem[{\citenamefont{Hardy and Towner}(2020)}]{Hardy:2020qwl}
\bibinfo{author}{\bibfnamefont{J.~C.} \bibnamefont{Hardy}} \bibnamefont{and}
  \bibinfo{author}{\bibfnamefont{I.~S.} \bibnamefont{Towner}},
  \bibinfo{journal}{Phys. Rev. C} \textbf{\bibinfo{volume}{102}},
  \bibinfo{pages}{045501} (\bibinfo{year}{2020}).

\bibitem[{\citenamefont{Gonzalez et~al.}(2021)}]{UCNt:2021pcg}
\bibinfo{author}{\bibfnamefont{F.~M.} \bibnamefont{Gonzalez}}
  \bibnamefont{et~al.} (\bibinfo{collaboration}{UCN\ensuremath{\tau}}),
  \bibinfo{journal}{Phys. Rev. Lett.} \textbf{\bibinfo{volume}{127}},
  \bibinfo{pages}{162501} (\bibinfo{year}{2021}), \eprint{2106.10375}.

\bibitem[{\citenamefont{M\"arkisch et~al.}(2019)}]{Markisch:2018ndu}
\bibinfo{author}{\bibfnamefont{B.}~\bibnamefont{M\"arkisch}}
  \bibnamefont{et~al.}, \bibinfo{journal}{Phys. Rev. Lett.}
  \textbf{\bibinfo{volume}{122}}, \bibinfo{pages}{242501}
  (\bibinfo{year}{2019}), \eprint{1812.04666}.

\bibitem[{\citenamefont{Beck et~al.}(2020)}]{Beck:2019xye}
\bibinfo{author}{\bibfnamefont{M.}~\bibnamefont{Beck}} \bibnamefont{et~al.},
  \bibinfo{journal}{Phys. Rev. C} \textbf{\bibinfo{volume}{101}},
  \bibinfo{pages}{055506} (\bibinfo{year}{2020}), \eprint{1908.04785}.

\bibitem[{\citenamefont{Cirigliano et~al.}(2003)\citenamefont{Cirigliano,
  Knecht, Neufeld, and Pichl}}]{Cirigliano:2002ng}
\bibinfo{author}{\bibfnamefont{V.}~\bibnamefont{Cirigliano}},
  \bibinfo{author}{\bibfnamefont{M.}~\bibnamefont{Knecht}},
  \bibinfo{author}{\bibfnamefont{H.}~\bibnamefont{Neufeld}}, \bibnamefont{and}
  \bibinfo{author}{\bibfnamefont{H.}~\bibnamefont{Pichl}},
  \bibinfo{journal}{Eur. Phys. J. C} \textbf{\bibinfo{volume}{27}},
  \bibinfo{pages}{255} (\bibinfo{year}{2003}), \eprint{hep-ph/0209226}.

\bibitem[{\citenamefont{Czarnecki et~al.}(2020)\citenamefont{Czarnecki,
  Marciano, and Sirlin}}]{Czarnecki:2019iwz}
\bibinfo{author}{\bibfnamefont{A.}~\bibnamefont{Czarnecki}},
  \bibinfo{author}{\bibfnamefont{W.~J.} \bibnamefont{Marciano}},
  \bibnamefont{and} \bibinfo{author}{\bibfnamefont{A.}~\bibnamefont{Sirlin}},
  \bibinfo{journal}{Phys. Rev. D} \textbf{\bibinfo{volume}{101}},
  \bibinfo{pages}{091301} (\bibinfo{year}{2020}), \eprint{1911.04685}.

\bibitem[{\citenamefont{Feng et~al.}(2020)\citenamefont{Feng, Gorchtein, Jin,
  Ma, and Seng}}]{Feng:2020zdc}
\bibinfo{author}{\bibfnamefont{X.}~\bibnamefont{Feng}},
  \bibinfo{author}{\bibfnamefont{M.}~\bibnamefont{Gorchtein}},
  \bibinfo{author}{\bibfnamefont{L.-C.} \bibnamefont{Jin}},
  \bibinfo{author}{\bibfnamefont{P.-X.} \bibnamefont{Ma}}, \bibnamefont{and}
  \bibinfo{author}{\bibfnamefont{C.-Y.} \bibnamefont{Seng}},
  \bibinfo{journal}{Phys. Rev. Lett.} \textbf{\bibinfo{volume}{124}},
  \bibinfo{pages}{192002} (\bibinfo{year}{2020}), \eprint{2003.09798}.

\bibitem[{\citenamefont{Po{\v c}ani{\'c} et~al.}(2004)}]{Pocanic:2003pf}
\bibinfo{author}{\bibfnamefont{D.}~\bibnamefont{Po{\v c}ani{\'c}}}
  \bibnamefont{et~al.}, \bibinfo{journal}{Phys. Rev. Lett.}
  \textbf{\bibinfo{volume}{93}}, \bibinfo{pages}{181803}
  (\bibinfo{year}{2004}), \eprint{hep-ex/0312030}.

\bibitem[{\citenamefont{Altmannshofer et~al.}(2022)}]{PIONEER:2022yag}
\bibinfo{author}{\bibfnamefont{W.}~\bibnamefont{Altmannshofer}}
  \bibnamefont{et~al.} (\bibinfo{collaboration}{PIONEER})
  (\bibinfo{year}{2022}), \eprint{2203.01981}.

\bibitem[{\citenamefont{Gorchtein and Seng}(2024)}]{Gorchtein:2023naa}
\bibinfo{author}{\bibfnamefont{M.}~\bibnamefont{Gorchtein}} \bibnamefont{and}
  \bibinfo{author}{\bibfnamefont{C.~Y.} \bibnamefont{Seng}},
  \bibinfo{journal}{Ann. Rev. Nucl. Part. Sci.} \textbf{\bibinfo{volume}{74}},
  \bibinfo{pages}{23} (\bibinfo{year}{2024}), \eprint{2311.00044}.

\bibitem[{\citenamefont{Sirlin}(1967)}]{Sirlin:1967zza}
\bibinfo{author}{\bibfnamefont{A.}~\bibnamefont{Sirlin}},
  \bibinfo{journal}{Phys. Rev.} \textbf{\bibinfo{volume}{164}},
  \bibinfo{pages}{1767} (\bibinfo{year}{1967}).

\bibitem[{\citenamefont{Jaus}(1972)}]{Jaus:1972hua}
\bibinfo{author}{\bibfnamefont{W.}~\bibnamefont{Jaus}}, \bibinfo{journal}{Phys.
  Lett. B} \textbf{\bibinfo{volume}{40}}, \bibinfo{pages}{616}
  (\bibinfo{year}{1972}).

\bibitem[{\citenamefont{Jaus and Rasche}(1987)}]{Jaus:1986te}
\bibinfo{author}{\bibfnamefont{W.}~\bibnamefont{Jaus}} \bibnamefont{and}
  \bibinfo{author}{\bibfnamefont{G.}~\bibnamefont{Rasche}},
  \bibinfo{journal}{Phys. Rev. D} \textbf{\bibinfo{volume}{35}},
  \bibinfo{pages}{3420} (\bibinfo{year}{1987}).

\bibitem[{\citenamefont{Sirlin and Zucchini}(1986)}]{Sirlin:1986cc}
\bibinfo{author}{\bibfnamefont{A.}~\bibnamefont{Sirlin}} \bibnamefont{and}
  \bibinfo{author}{\bibfnamefont{R.}~\bibnamefont{Zucchini}},
  \bibinfo{journal}{Phys. Rev. Lett.} \textbf{\bibinfo{volume}{57}},
  \bibinfo{pages}{1994} (\bibinfo{year}{1986}).

\bibitem[{\citenamefont{Gorchtein}(2019)}]{Gorchtein:2018fxl}
\bibinfo{author}{\bibfnamefont{M.}~\bibnamefont{Gorchtein}},
  \bibinfo{journal}{Phys. Rev. Lett.} \textbf{\bibinfo{volume}{123}},
  \bibinfo{pages}{042503} (\bibinfo{year}{2019}), \eprint{1812.04229}.

\bibitem[{\citenamefont{Seng and Gorchtein}(2023{\natexlab{a}})}]{Seng:2022cnq}
\bibinfo{author}{\bibfnamefont{C.-Y.} \bibnamefont{Seng}} \bibnamefont{and}
  \bibinfo{author}{\bibfnamefont{M.}~\bibnamefont{Gorchtein}},
  \bibinfo{journal}{Phys. Rev. C} \textbf{\bibinfo{volume}{107}},
  \bibinfo{pages}{035503} (\bibinfo{year}{2023}{\natexlab{a}}),
  \eprint{2211.10214}.

\bibitem[{\citenamefont{Miller and Schwenk}(2008)}]{Miller:2008my}
\bibinfo{author}{\bibfnamefont{G.~A.} \bibnamefont{Miller}} \bibnamefont{and}
  \bibinfo{author}{\bibfnamefont{A.}~\bibnamefont{Schwenk}},
  \bibinfo{journal}{Phys. Rev. C} \textbf{\bibinfo{volume}{78}},
  \bibinfo{pages}{035501} (\bibinfo{year}{2008}), \eprint{0805.0603}.

\bibitem[{\citenamefont{Miller and Schwenk}(2009)}]{Miller:2009cg}
\bibinfo{author}{\bibfnamefont{G.~A.} \bibnamefont{Miller}} \bibnamefont{and}
  \bibinfo{author}{\bibfnamefont{A.}~\bibnamefont{Schwenk}},
  \bibinfo{journal}{Phys. Rev. C} \textbf{\bibinfo{volume}{80}},
  \bibinfo{pages}{064319} (\bibinfo{year}{2009}), \eprint{0910.2790}.

\bibitem[{\citenamefont{Martin et~al.}(2021)\citenamefont{Martin, Stroberg,
  Holt, and Leach}}]{Martin:2021bud}
\bibinfo{author}{\bibfnamefont{M.~S.} \bibnamefont{Martin}},
  \bibinfo{author}{\bibfnamefont{S.~R.} \bibnamefont{Stroberg}},
  \bibinfo{author}{\bibfnamefont{J.~D.} \bibnamefont{Holt}}, \bibnamefont{and}
  \bibinfo{author}{\bibfnamefont{K.~G.} \bibnamefont{Leach}},
  \bibinfo{journal}{Phys. Rev. C} \textbf{\bibinfo{volume}{104}},
  \bibinfo{pages}{014324} (\bibinfo{year}{2021}), \eprint{2101.11826}.

\bibitem[{\citenamefont{Condren and Miller}(2022)}]{Condren:2022dji}
\bibinfo{author}{\bibfnamefont{L.}~\bibnamefont{Condren}} \bibnamefont{and}
  \bibinfo{author}{\bibfnamefont{G.~A.} \bibnamefont{Miller}},
  \bibinfo{journal}{Phys. Rev. C} \textbf{\bibinfo{volume}{106}},
  \bibinfo{pages}{L062501} (\bibinfo{year}{2022}), \eprint{2201.10651}.

\bibitem[{\citenamefont{Seng and Gorchtein}(2023{\natexlab{b}})}]{Seng:2022epj}
\bibinfo{author}{\bibfnamefont{C.-Y.} \bibnamefont{Seng}} \bibnamefont{and}
  \bibinfo{author}{\bibfnamefont{M.}~\bibnamefont{Gorchtein}},
  \bibinfo{journal}{Phys. Lett. B} \textbf{\bibinfo{volume}{838}},
  \bibinfo{pages}{137654} (\bibinfo{year}{2023}{\natexlab{b}}),
  \eprint{2208.03037}.

\bibitem[{\citenamefont{Crawford and Miller}(2022)}]{Crawford:2022yhi}
\bibinfo{author}{\bibfnamefont{J.~W.} \bibnamefont{Crawford}} \bibnamefont{and}
  \bibinfo{author}{\bibfnamefont{G.~A.} \bibnamefont{Miller}},
  \bibinfo{journal}{Phys. Rev. C} \textbf{\bibinfo{volume}{106}},
  \bibinfo{pages}{065502} (\bibinfo{year}{2022}), \eprint{2209.10603}.

\bibitem[{\citenamefont{Seng and Gorchtein}(2024{\natexlab{a}})}]{Seng:2023cvt}
\bibinfo{author}{\bibfnamefont{C.-Y.} \bibnamefont{Seng}} \bibnamefont{and}
  \bibinfo{author}{\bibfnamefont{M.}~\bibnamefont{Gorchtein}},
  \bibinfo{journal}{Phys. Rev. C} \textbf{\bibinfo{volume}{109}},
  \bibinfo{pages}{044302} (\bibinfo{year}{2024}{\natexlab{a}}),
  \eprint{2304.03800}.

\bibitem[{\citenamefont{Seng and Gorchtein}(2024{\natexlab{b}})}]{Seng:2023cgl}
\bibinfo{author}{\bibfnamefont{C.-Y.} \bibnamefont{Seng}} \bibnamefont{and}
  \bibinfo{author}{\bibfnamefont{M.}~\bibnamefont{Gorchtein}},
  \bibinfo{journal}{Phys. Rev. C} \textbf{\bibinfo{volume}{109}},
  \bibinfo{pages}{045501} (\bibinfo{year}{2024}{\natexlab{b}}),
  \eprint{2309.16893}.

\bibitem[{\citenamefont{Cirigliano
  et~al.}(2023{\natexlab{b}})\citenamefont{Cirigliano, Dekens, Mereghetti, and
  Tomalak}}]{Cirigliano:2023fnz}
\bibinfo{author}{\bibfnamefont{V.}~\bibnamefont{Cirigliano}},
  \bibinfo{author}{\bibfnamefont{W.}~\bibnamefont{Dekens}},
  \bibinfo{author}{\bibfnamefont{E.}~\bibnamefont{Mereghetti}},
  \bibnamefont{and} \bibinfo{author}{\bibfnamefont{O.}~\bibnamefont{Tomalak}},
  \bibinfo{journal}{Phys. Rev. D} \textbf{\bibinfo{volume}{108}},
  \bibinfo{pages}{053003} (\bibinfo{year}{2023}{\natexlab{b}}),
  \eprint{2306.03138}.

\bibitem[{\citenamefont{Ma et~al.}(2024)\citenamefont{Ma, Feng, Gorchtein, Jin,
  Liu, Seng, Wang, and Zhang}}]{Ma:2023kfr}
\bibinfo{author}{\bibfnamefont{P.-X.} \bibnamefont{Ma}},
  \bibinfo{author}{\bibfnamefont{X.}~\bibnamefont{Feng}},
  \bibinfo{author}{\bibfnamefont{M.}~\bibnamefont{Gorchtein}},
  \bibinfo{author}{\bibfnamefont{L.-C.} \bibnamefont{Jin}},
  \bibinfo{author}{\bibfnamefont{K.-F.} \bibnamefont{Liu}},
  \bibinfo{author}{\bibfnamefont{C.-Y.} \bibnamefont{Seng}},
  \bibinfo{author}{\bibfnamefont{B.-G.} \bibnamefont{Wang}}, \bibnamefont{and}
  \bibinfo{author}{\bibfnamefont{Z.-L.} \bibnamefont{Zhang}},
  \bibinfo{journal}{Phys. Rev. Lett.} \textbf{\bibinfo{volume}{132}},
  \bibinfo{pages}{191901} (\bibinfo{year}{2024}), \eprint{2308.16755}.

\bibitem[{\citenamefont{Gennari et~al.}(2024)\citenamefont{Gennari, Drissi,
  Gorchtein, Navratil, and Seng}}]{Gennari:2024sbn}
\bibinfo{author}{\bibfnamefont{M.}~\bibnamefont{Gennari}},
  \bibinfo{author}{\bibfnamefont{M.}~\bibnamefont{Drissi}},
  \bibinfo{author}{\bibfnamefont{M.}~\bibnamefont{Gorchtein}},
  \bibinfo{author}{\bibfnamefont{P.}~\bibnamefont{Navratil}}, \bibnamefont{and}
  \bibinfo{author}{\bibfnamefont{C.-Y.} \bibnamefont{Seng}}
  (\bibinfo{year}{2024}), \eprint{2405.19281}.

\bibitem[{\citenamefont{Cirigliano et~al.}(2018)\citenamefont{Cirigliano,
  Dekens, de~Vries, Graesser, Mereghetti, Pastore, and
  Van~Kolck}}]{Cirigliano:2018hja}
\bibinfo{author}{\bibfnamefont{V.}~\bibnamefont{Cirigliano}},
  \bibinfo{author}{\bibfnamefont{W.}~\bibnamefont{Dekens}},
  \bibinfo{author}{\bibfnamefont{J.}~\bibnamefont{de~Vries}},
  \bibinfo{author}{\bibfnamefont{M.~L.} \bibnamefont{Graesser}},
  \bibinfo{author}{\bibfnamefont{E.}~\bibnamefont{Mereghetti}},
  \bibinfo{author}{\bibfnamefont{S.}~\bibnamefont{Pastore}}, \bibnamefont{and}
  \bibinfo{author}{\bibfnamefont{U.}~\bibnamefont{Van~Kolck}},
  \bibinfo{journal}{Phys. Rev. Lett.} \textbf{\bibinfo{volume}{120}},
  \bibinfo{pages}{202001} (\bibinfo{year}{2018}), \eprint{1802.10097}.

\bibitem[{\citenamefont{Cirigliano et~al.}(2019)\citenamefont{Cirigliano,
  Dekens, de~Vries, Graesser, Mereghetti, Pastore, Piarulli, Van~Kolck, and
  Wiringa}}]{Cirigliano:2019vdj}
\bibinfo{author}{\bibfnamefont{V.}~\bibnamefont{Cirigliano}},
  \bibinfo{author}{\bibfnamefont{W.}~\bibnamefont{Dekens}},
  \bibinfo{author}{\bibfnamefont{J.}~\bibnamefont{de~Vries}},
  \bibinfo{author}{\bibfnamefont{M.~L.} \bibnamefont{Graesser}},
  \bibinfo{author}{\bibfnamefont{E.}~\bibnamefont{Mereghetti}},
  \bibinfo{author}{\bibfnamefont{S.}~\bibnamefont{Pastore}},
  \bibinfo{author}{\bibfnamefont{M.}~\bibnamefont{Piarulli}},
  \bibinfo{author}{\bibfnamefont{U.}~\bibnamefont{Van~Kolck}},
  \bibnamefont{and} \bibinfo{author}{\bibfnamefont{R.~B.}
  \bibnamefont{Wiringa}}, \bibinfo{journal}{Phys. Rev. C}
  \textbf{\bibinfo{volume}{100}}, \bibinfo{pages}{055504}
  (\bibinfo{year}{2019}), \eprint{1907.11254}.

\bibitem[{\citenamefont{Cirigliano
  et~al.}(2021{\natexlab{a}})\citenamefont{Cirigliano, Dekens, de~Vries,
  Hoferichter, and Mereghetti}}]{Cirigliano:2020dmx}
\bibinfo{author}{\bibfnamefont{V.}~\bibnamefont{Cirigliano}},
  \bibinfo{author}{\bibfnamefont{W.}~\bibnamefont{Dekens}},
  \bibinfo{author}{\bibfnamefont{J.}~\bibnamefont{de~Vries}},
  \bibinfo{author}{\bibfnamefont{M.}~\bibnamefont{Hoferichter}},
  \bibnamefont{and}
  \bibinfo{author}{\bibfnamefont{E.}~\bibnamefont{Mereghetti}},
  \bibinfo{journal}{Phys. Rev. Lett.} \textbf{\bibinfo{volume}{126}},
  \bibinfo{pages}{172002} (\bibinfo{year}{2021}{\natexlab{a}}),
  \eprint{2012.11602}.

\bibitem[{\citenamefont{Cirigliano
  et~al.}(2021{\natexlab{b}})\citenamefont{Cirigliano, Dekens, de~Vries,
  Hoferichter, and Mereghetti}}]{Cirigliano:2021qko}
\bibinfo{author}{\bibfnamefont{V.}~\bibnamefont{Cirigliano}},
  \bibinfo{author}{\bibfnamefont{W.}~\bibnamefont{Dekens}},
  \bibinfo{author}{\bibfnamefont{J.}~\bibnamefont{de~Vries}},
  \bibinfo{author}{\bibfnamefont{M.}~\bibnamefont{Hoferichter}},
  \bibnamefont{and}
  \bibinfo{author}{\bibfnamefont{E.}~\bibnamefont{Mereghetti}},
  \bibinfo{journal}{JHEP} \textbf{\bibinfo{volume}{05}}, \bibinfo{pages}{289}
  (\bibinfo{year}{2021}{\natexlab{b}}), \eprint{2102.03371}.

\bibitem[{\citenamefont{Wirth et~al.}(2021)\citenamefont{Wirth, Yao, and
  Hergert}}]{Wirth:2021pij}
\bibinfo{author}{\bibfnamefont{R.}~\bibnamefont{Wirth}},
  \bibinfo{author}{\bibfnamefont{J.~M.} \bibnamefont{Yao}}, \bibnamefont{and}
  \bibinfo{author}{\bibfnamefont{H.}~\bibnamefont{Hergert}},
  \bibinfo{journal}{Phys. Rev. Lett.} \textbf{\bibinfo{volume}{127}},
  \bibinfo{pages}{242502} (\bibinfo{year}{2021}), \eprint{2105.05415}.

\bibitem[{\citenamefont{Jokiniemi et~al.}(2021)\citenamefont{Jokiniemi,
  Soriano, and Men\'endez}}]{Jokiniemi:2021qqv}
\bibinfo{author}{\bibfnamefont{L.}~\bibnamefont{Jokiniemi}},
  \bibinfo{author}{\bibfnamefont{P.}~\bibnamefont{Soriano}}, \bibnamefont{and}
  \bibinfo{author}{\bibfnamefont{J.}~\bibnamefont{Men\'endez}},
  \bibinfo{journal}{Phys. Lett. B} \textbf{\bibinfo{volume}{823}},
  \bibinfo{pages}{136720} (\bibinfo{year}{2021}), \eprint{2107.13354}.

\bibitem[{\citenamefont{Belley et~al.}(2024)}]{Belley:2023lec}
\bibinfo{author}{\bibfnamefont{A.}~\bibnamefont{Belley}} \bibnamefont{et~al.},
  \bibinfo{journal}{Phys. Rev. Lett.} \textbf{\bibinfo{volume}{132}},
  \bibinfo{pages}{182502} (\bibinfo{year}{2024}), \eprint{2308.15634}.

\bibitem[{\citenamefont{Cirigliano
  et~al.}(2024{\natexlab{b}})\citenamefont{Cirigliano, Dekens, de~Vries,
  Gandolfi, Hoferichter, and Mereghetti}}]{Cirigliano:2024msg}
\bibinfo{author}{\bibfnamefont{V.}~\bibnamefont{Cirigliano}},
  \bibinfo{author}{\bibfnamefont{W.}~\bibnamefont{Dekens}},
  \bibinfo{author}{\bibfnamefont{J.}~\bibnamefont{de~Vries}},
  \bibinfo{author}{\bibfnamefont{S.}~\bibnamefont{Gandolfi}},
  \bibinfo{author}{\bibfnamefont{M.}~\bibnamefont{Hoferichter}},
  \bibnamefont{and}
  \bibinfo{author}{\bibfnamefont{E.}~\bibnamefont{Mereghetti}},
  \bibinfo{journal}{Phys. Rev. C} \textbf{\bibinfo{volume}{110}},
  \bibinfo{pages}{055502} (\bibinfo{year}{2024}{\natexlab{b}}),
  \eprint{2405.18464}.

\bibitem[{\citenamefont{Cirigliano
  et~al.}(2022{\natexlab{b}})\citenamefont{Cirigliano, de~Vries, Hayen,
  Mereghetti, and Walker-Loud}}]{Cirigliano:2022hob}
\bibinfo{author}{\bibfnamefont{V.}~\bibnamefont{Cirigliano}},
  \bibinfo{author}{\bibfnamefont{J.}~\bibnamefont{de~Vries}},
  \bibinfo{author}{\bibfnamefont{L.}~\bibnamefont{Hayen}},
  \bibinfo{author}{\bibfnamefont{E.}~\bibnamefont{Mereghetti}},
  \bibnamefont{and}
  \bibinfo{author}{\bibfnamefont{A.}~\bibnamefont{Walker-Loud}},
  \bibinfo{journal}{Phys. Rev. Lett.} \textbf{\bibinfo{volume}{129}},
  \bibinfo{pages}{121801} (\bibinfo{year}{2022}{\natexlab{b}}),
  \eprint{2202.10439}.

\bibitem[{\citenamefont{Jenkins and Manohar}(1991)}]{Jenkins:1990jv}
\bibinfo{author}{\bibfnamefont{E.~E.} \bibnamefont{Jenkins}} \bibnamefont{and}
  \bibinfo{author}{\bibfnamefont{A.~V.} \bibnamefont{Manohar}},
  \bibinfo{journal}{Phys. Lett. B} \textbf{\bibinfo{volume}{255}},
  \bibinfo{pages}{558} (\bibinfo{year}{1991}).

\bibitem[{\citenamefont{Bernard et~al.}(1995)\citenamefont{Bernard, Kaiser, and
  Mei{\ss}ner}}]{Bernard:1995dp}
\bibinfo{author}{\bibfnamefont{V.}~\bibnamefont{Bernard}},
  \bibinfo{author}{\bibfnamefont{N.}~\bibnamefont{Kaiser}}, \bibnamefont{and}
  \bibinfo{author}{\bibfnamefont{U.-G.} \bibnamefont{Mei{\ss}ner}},
  \bibinfo{journal}{Int. J. Mod. Phys. E} \textbf{\bibinfo{volume}{4}},
  \bibinfo{pages}{193} (\bibinfo{year}{1995}), \eprint{hep-ph/9501384}.

\bibitem[{\citenamefont{Beneke and Smirnov}(1998)}]{Beneke:1997zp}
\bibinfo{author}{\bibfnamefont{M.}~\bibnamefont{Beneke}} \bibnamefont{and}
  \bibinfo{author}{\bibfnamefont{V.~A.} \bibnamefont{Smirnov}},
  \bibinfo{journal}{Nucl. Phys. B} \textbf{\bibinfo{volume}{522}},
  \bibinfo{pages}{321} (\bibinfo{year}{1998}), \eprint{hep-ph/9711391}.

\bibitem[{\citenamefont{Smirnov}(2002)}]{Smirnov:2002pj}
\bibinfo{author}{\bibfnamefont{V.~A.} \bibnamefont{Smirnov}},
  \bibinfo{journal}{Springer Tracts Mod. Phys.} \textbf{\bibinfo{volume}{177}},
  \bibinfo{pages}{1} (\bibinfo{year}{2002}).

\bibitem[{\citenamefont{Fermi}(1934)}]{Fermi:1934sk}
\bibinfo{author}{\bibfnamefont{E.}~\bibnamefont{Fermi}},
  \bibinfo{journal}{Nuovo Cim.} \textbf{\bibinfo{volume}{11}},
  \bibinfo{pages}{1} (\bibinfo{year}{1934}).

\bibitem[{\citenamefont{Jaus and Rasche}(1970)}]{Jaus:1970tah}
\bibinfo{author}{\bibfnamefont{W.}~\bibnamefont{Jaus}} \bibnamefont{and}
  \bibinfo{author}{\bibfnamefont{G.}~\bibnamefont{Rasche}},
  \bibinfo{journal}{Nucl. Phys. A} \textbf{\bibinfo{volume}{143}},
  \bibinfo{pages}{202} (\bibinfo{year}{1970}).

\bibitem[{\citenamefont{Sirlin}(1978)}]{Sirlin:1977sv}
\bibinfo{author}{\bibfnamefont{A.}~\bibnamefont{Sirlin}},
  \bibinfo{journal}{Rev. Mod. Phys.} \textbf{\bibinfo{volume}{50}},
  \bibinfo{pages}{573} (\bibinfo{year}{1978}), \bibinfo{note}{[Erratum: Rev.
  Mod. Phys. {\bf 50}, 905 (1978)]}.

\bibitem[{\citenamefont{Sirlin}(1987)}]{Sirlin:1986hpu}
\bibinfo{author}{\bibfnamefont{A.}~\bibnamefont{Sirlin}},
  \bibinfo{journal}{Phys. Rev. D} \textbf{\bibinfo{volume}{35}},
  \bibinfo{pages}{3423} (\bibinfo{year}{1987}).

\bibitem[{\citenamefont{Ando et~al.}(2004)\citenamefont{Ando, Fearing, Gudkov,
  Kubodera, Myhrer, Nakamura, and Sato}}]{Ando:2004rk}
\bibinfo{author}{\bibfnamefont{S.}~\bibnamefont{Ando}},
  \bibinfo{author}{\bibfnamefont{H.~W.} \bibnamefont{Fearing}},
  \bibinfo{author}{\bibfnamefont{V.~P.} \bibnamefont{Gudkov}},
  \bibinfo{author}{\bibfnamefont{K.}~\bibnamefont{Kubodera}},
  \bibinfo{author}{\bibfnamefont{F.}~\bibnamefont{Myhrer}},
  \bibinfo{author}{\bibfnamefont{S.}~\bibnamefont{Nakamura}}, \bibnamefont{and}
  \bibinfo{author}{\bibfnamefont{T.}~\bibnamefont{Sato}},
  \bibinfo{journal}{Phys. Lett. B} \textbf{\bibinfo{volume}{595}},
  \bibinfo{pages}{250} (\bibinfo{year}{2004}), \eprint{nucl-th/0402100}.

\bibitem[{\citenamefont{Hill and Plestid}(2024{\natexlab{a}})}]{Hill:2023acw}
\bibinfo{author}{\bibfnamefont{R.~J.} \bibnamefont{Hill}} \bibnamefont{and}
  \bibinfo{author}{\bibfnamefont{R.}~\bibnamefont{Plestid}},
  \bibinfo{journal}{Phys. Rev. Lett.} \textbf{\bibinfo{volume}{133}},
  \bibinfo{pages}{021803} (\bibinfo{year}{2024}{\natexlab{a}}),
  \eprint{2309.07343}.

\bibitem[{\citenamefont{Hill and Plestid}(2024{\natexlab{b}})}]{Hill:2023bfh}
\bibinfo{author}{\bibfnamefont{R.~J.} \bibnamefont{Hill}} \bibnamefont{and}
  \bibinfo{author}{\bibfnamefont{R.}~\bibnamefont{Plestid}},
  \bibinfo{journal}{Phys. Rev. D} \textbf{\bibinfo{volume}{109}},
  \bibinfo{pages}{056006} (\bibinfo{year}{2024}{\natexlab{b}}),
  \eprint{2309.15929}.

\bibitem[{\citenamefont{Borah et~al.}(2024)\citenamefont{Borah, Hill, and
  Plestid}}]{Borah:2024ghn}
\bibinfo{author}{\bibfnamefont{K.}~\bibnamefont{Borah}},
  \bibinfo{author}{\bibfnamefont{R.~J.} \bibnamefont{Hill}}, \bibnamefont{and}
  \bibinfo{author}{\bibfnamefont{R.}~\bibnamefont{Plestid}},
  \bibinfo{journal}{Phys. Rev. D} \textbf{\bibinfo{volume}{109}},
  \bibinfo{pages}{113007} (\bibinfo{year}{2024}), \eprint{2402.13307}.

\bibitem[{\citenamefont{Pav\'on~Valderrama and
  Phillips}(2015)}]{PavonValderrama:2014zeq}
\bibinfo{author}{\bibfnamefont{M.}~\bibnamefont{Pav\'on~Valderrama}}
  \bibnamefont{and} \bibinfo{author}{\bibfnamefont{D.~R.}
  \bibnamefont{Phillips}}, \bibinfo{journal}{Phys. Rev. Lett.}
  \textbf{\bibinfo{volume}{114}}, \bibinfo{pages}{082502}
  (\bibinfo{year}{2015}), \eprint{1407.0437}.

\bibitem[{\citenamefont{Towner}(1992)}]{Towner:1992xm}
\bibinfo{author}{\bibfnamefont{I.~S.} \bibnamefont{Towner}},
  \bibinfo{journal}{Nucl. Phys. A} \textbf{\bibinfo{volume}{540}},
  \bibinfo{pages}{478} (\bibinfo{year}{1992}).

\end{thebibliography}

\end{document}